\documentclass[pdftex]{article}
\pdfoutput=1
\usepackage{amssymb}
\usepackage{icrctc07}
\title{The end of the Galactic spectrum}

\shorttitle{The end of the Galactic spectrum}

\authors{C. De Donato$^{1}$ , G. A. Medina-Tanco$^{2}$}

\shortauthors{De Donato C. and Medina-Tanco G.}

\afiliations{
$^1$Dipartimento di Fisica dell'Universit\`a degli Studi di Milano and INFN, Milano, Italy, \\
$^2$Dep. Altas Energias, Inst. de Ciencias Nucleares, Universidad
Nacional Autonoma de Mexico, Mexico DF, CP 04510}
\email{cinzia.dedonato@mi.infn.it}

\abstract{We use a diffusion galactic model to analyze the end of
the Galactic cosmic ray spectrum and its mixing with the
extragalactic cosmic ray flux. We analyze the transition between
Galactic and extragalactic components using two different
extragalactic models. We compare the sum of the diffusive galactic
spectrum and extragalactic spectrum  with the available experimental
data.}

\begin{document}
\maketitle

\section{Introduction}
The cosmic ray energy spectrum extends for many orders of magnitude
with a power law  index $\approx2.7$. Along this range of energies,
the three spectral features are known: the first knee at $E\approx
3~PeV$, the second knee at $E\approx0.5~EeV$ and the ankle, a dip
extending from the second knee to beyond $10~EeV$. The nature of the
second knee and of the ankle are still uncertain \cite{GMTanco2006};
a possible interpretation of the two features is the transition
between the galactic and extragalactic components. At energies
between $10^{17}-10^{18}~eV$ the galactic supernova remnants (SNRs)
 are expected to become inefficient as accelerators. This
fact, combined with magnetic deconfinement should mark the end of
the Galactic component of cosmic rays, although the picture could be
confused by the existence of additional Galactic accelerators at
higher energy. On the other hand, at energies above the second knee,
extragalactic particles are able to travel from the nearest
extragalactic sources in less than a Hubble time. Consequently, the
spectrum may present above $10^{17.5}~eV$ a growing extragalactic
component that becomes dominant above $10^{19}~eV$. The region
between the second knee and the ankle could be the transition region
between the galactic and extragalactic components.\\
In this work we analyze the transition region comparing the diffusive galactic
spectrum from SNRs with two different models of extragalactic
spectrum, one in which only protons \cite{Berezinsky2006} are
injected at the sources and another in which a mixed composition
containing heavy nuclei \cite{Allard2007} is injected.

\section{Diffusion Galactic model}

We used the numerical diffusive propagation code
GALPROP \cite{Strong1998,Strong2001} to reproduce the galactic
spectrum from SuperNova Remnants (SNRs). The diffusive model is 
axisymmetric. The propagation region is, in cylindrical coordinates,
bounded by $R=R_h=30~kpc$ and $z=z_h=4~kpc$, beyond which free
escape is assumed. The propagation equation is:

\vskip -0.3cm

\begin{eqnarray}\label{diffeq}
\frac{\partial\psi}{\partial t} & = & q(\vec{r},p)+ \vec{\nabla}
\cdot (D_{xx}\vec{\nabla} \psi)+ \nonumber \\
& & -\frac{\partial}{\partial
p}(\dot{p}\psi)-\frac{1}{\tau_f}\psi-\frac{1}{\tau_r}\psi
\end{eqnarray}

where $\psi(\vec{r},p,t)$ is the density per unit of total particle
momentum, $q(\vec{r},p)$ is the source term, $D_{xx}$ is the spatial
diffusion coefficient, $\dot{p}= dp/dt$ is the momentum loss rate
and $\tau_f$ and $\tau_r$ are the time scale of fragmentation and
the time scale of radioactive decay respectively. The diffusion
coefficient is taken as $\beta D_0(\rho / \rho_D)^{\delta}$ , where
$\rho$ is the particle rigidity, $D_0$ is the diffusion coefficient at a reference rigidity $\rho_D$ and $\delta=0.6$. 
The distribution of cosmic rays sources used is
that of Galactic SNRs (in agreement with EGRET gamma-ray data)
\cite{Strong1998}. The injection spectrum is a a power law function
in rigidity with a break at rigidity $\rho_0$, beyond which it falls
exponentially with a rigidity scale $\rho_c$:

\begin{displaymath}
I(\rho)= \left\{ \begin{array}{ll}
\label{InjSp}
 \left(\frac{\rho}{\rho_0}\right)^{-\alpha} & \rho\leq\rho_0\\
\exp\left[-(\frac{\rho}{\rho_0}-1)/\frac{\rho_c}{\rho_0})\right] & \rho>\rho_0 \nonumber
\end{array}\right.
\end{displaymath}

where $\alpha=2.05$, $\rho_0=1.8~PV$ and $\rho_c=1.26~PV$. Stable
nuclei with $Z<26$ are injected, with energy independent isotopic
abundances derived from low energy CR measurements\cite{Strong2001}.

Detailed and realistic interstellar molecular (H$_2$), atomic (H)
and ionized (HI) hydrogen distributions are used
\cite{Moskalenko2002}.

\section{Diffusive Galactic spectrum}\label{Diffusive Galactic spectrum}

\begin{figure}
\begin{center}
\includegraphics [width=0.48\textwidth]{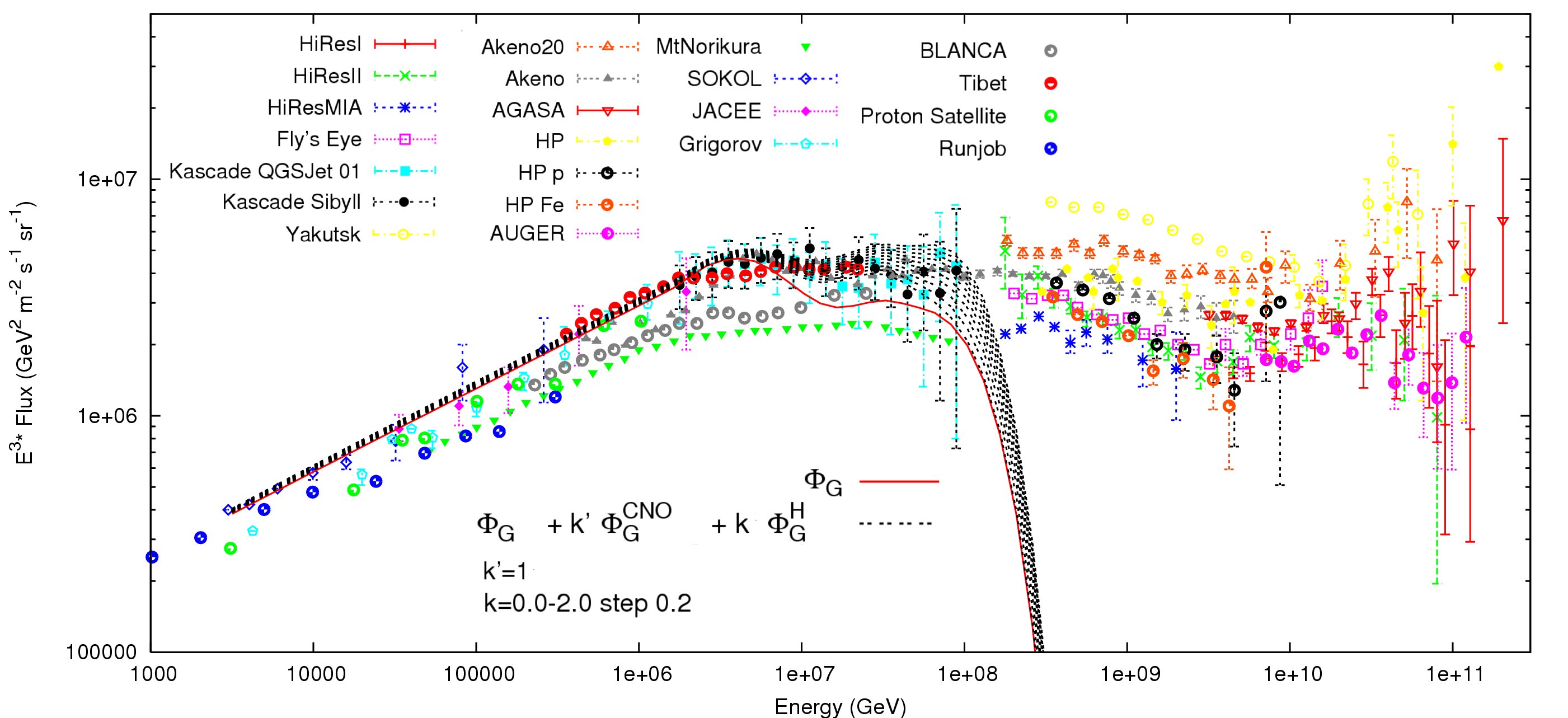}
\end{center}
\caption{Diffusive Galactic spectrum,  $\Phi_G$, with
renormalization of the CNO group (Z:6-12), $\Phi_G^{CNO}$ and of the
heavy component ($Z:19-26$), $\Phi_G^{H}$. Agreement with
experimental data is obtained for $k^\prime \approx 1$ and 
$k\approx0.8-1.2$.}\label{GALrenormC}
\end{figure}

The diffusive galactic spectrum has been normalized to match KASCADE
data at $\sim 3 \times 10^6$ GeV. While with this renormalization
our spectrum agrees with JACEE and Sokol data at lower energies,
beyond the knee the diffusive spectrum presents a strong deficit of
flux. Since at $E> 10^7$~GeV the composition is dominated by
intermediate ($Z:6-12$) and heavier ($Z:19-26$) nuclei, we
renormalize these components by a factor of $2$, which produces a
good agreement with the experimental data (see Fig.
\ref{GALrenormC}).

The renormalized diffusive galactic spectrum reproduces well the
data up to $E\approx10^{8}~GeV$, beyond which the spectrum falls
steeply because of end of the SNR acceleration efficiency.

\section{Extragalactic spectrum}\label{EG}

In order to study how the transition between the galactic and
extragalactic components takes place, we compare the galactic
spectrum originated in SNR with two different possible scenarios for
the extragalactic component.

In the first model \cite{Berezinsky2006}, a pure proton
extragalactic spectrum, accelerated by a homogeneous distribution of
cosmic sources, is considered. Local overdensities/deficits  of
UHECR sources affect the shape of the GZK modulation, but do not
affect the low energy region where the matching with the Galactic
spectrum occurs. 
Different cases of local overdensity/deficity of sources are considered within this model.

In the second model \cite{Allard2007}, the extragalactic spectrum is
calculated for a mixed composition at injection typical of low
energy cosmic rays for  different source evolution models in red shift.

In both cases, the various parameters of the models are tuned to fit
the available CR data at UHE and are, in that highest energy regime,
experimentally indistinguishable at present.

\section{Combined spectrum: matching Galactic and extragalactic components}

In order to study how the transition between the Galactic and
extragalactic components takes place, we subtract the combined
theoretical (Galactic plus extragalactic) spectrum from the
available data. Two different approaches are used.

\begin{figure}
\begin{center}
\includegraphics [width=0.48\textwidth]{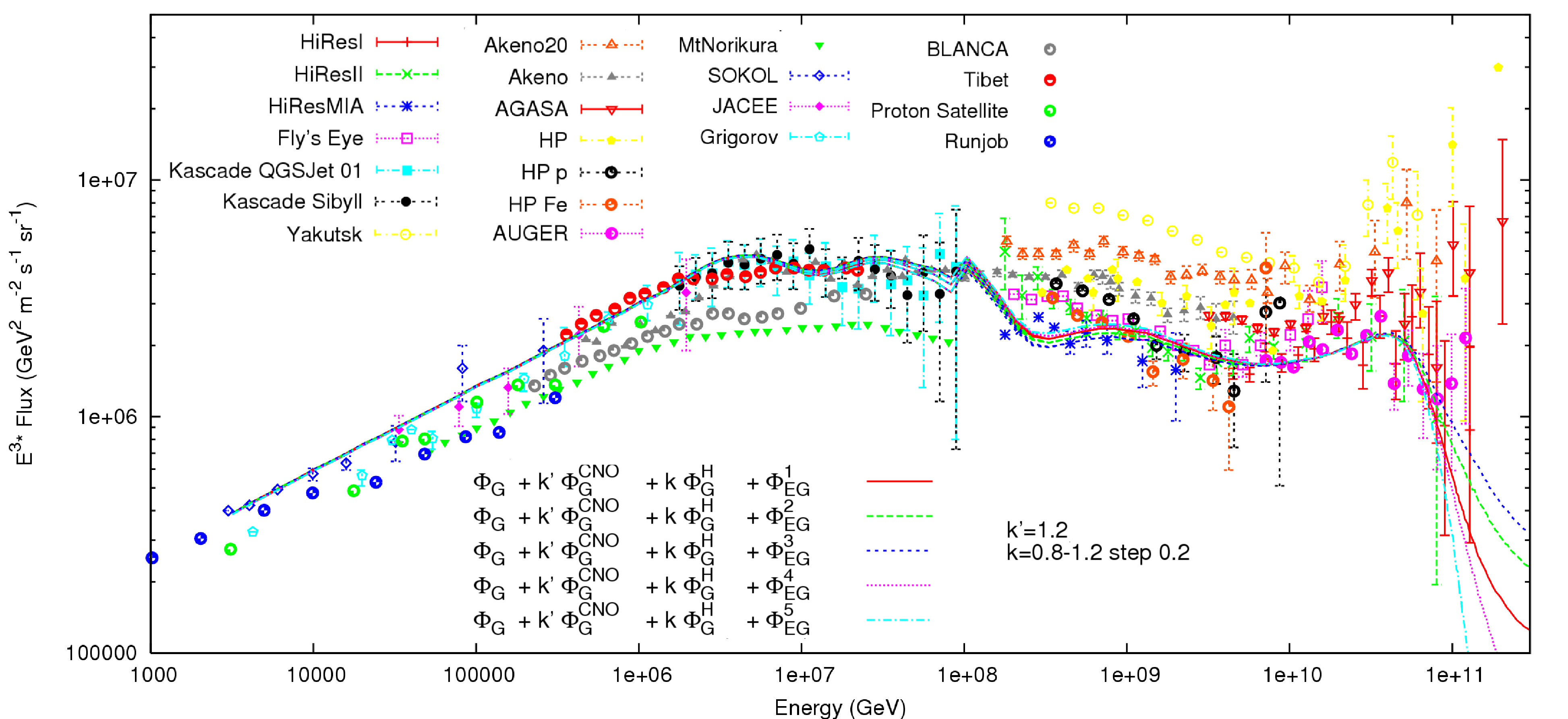}
\end{center}
\caption{Galactive and Extragalactic spectrum matching for the {\it
proton\/} models for a EG lower energy limit of $10^8~GeV$. The sum
of the renormalized diffusive Galactic spectrum and different
extragalactic spectrum models ($\Phi_{EG}$) is
shown for different renormalizations of the heavy component
($\Phi_G^{H}$). The CNO group ($\Phi_G^{CNO}$) of the diffusive
Galactic spectrum ($\Phi_G$) has been renormalized by a factor
$2.2$. Different cases of local overdensity/deficit of sources are considered \cite{Berezinsky2006}: 
(1) universal spectrum, (2) and (3) overdensity of sources, (4) and (5) deficit of sources .}\label{BeGR}
\end{figure}

\begin{figure}
\begin{center}
\includegraphics [width=0.48\textwidth]{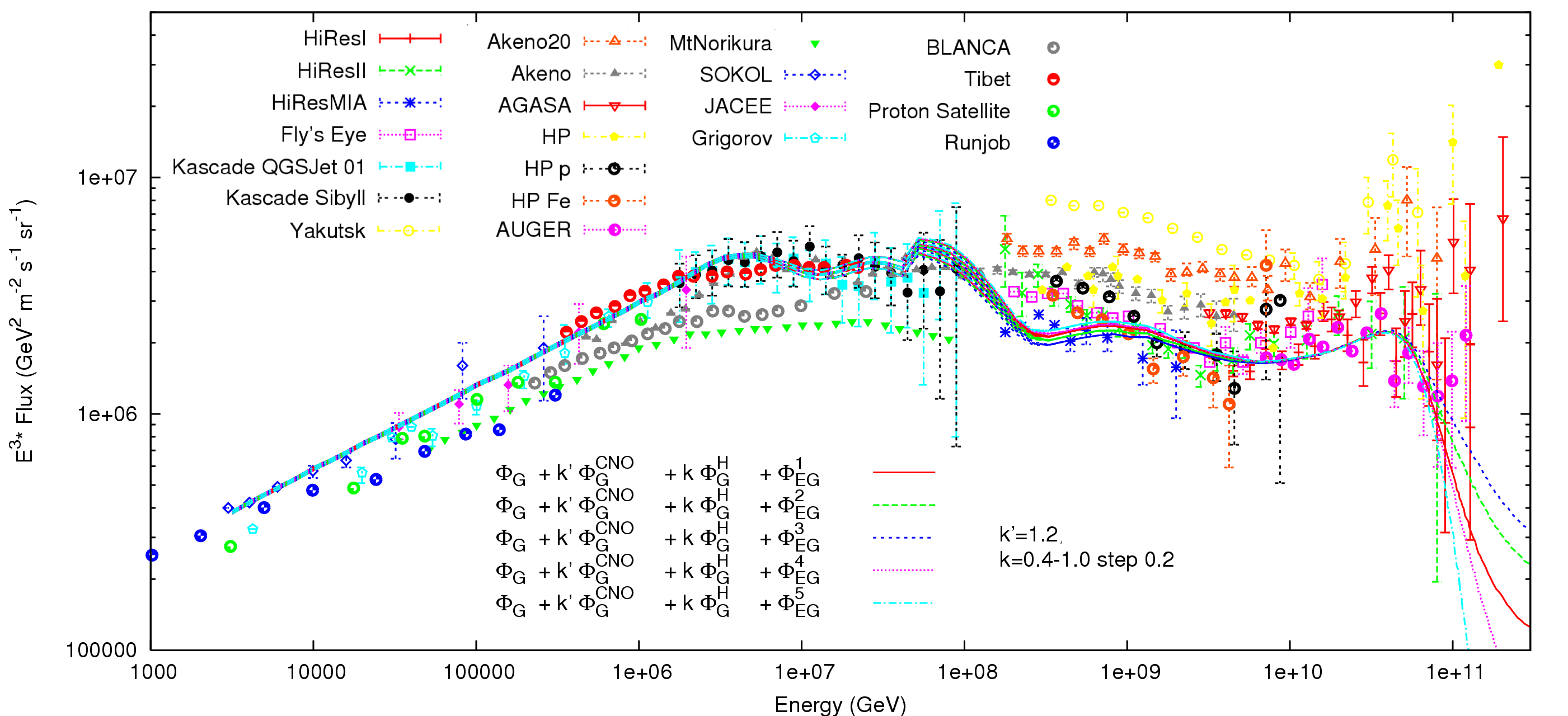}
\end{center}
\caption{Idem to Fig. \ref{BeGR} but for the {\it proton\/} model
with a lower energy limit of $5*10^7$~GeV.}\label{BeGRLL}
\end{figure}

\begin{figure}
\begin{center}
\includegraphics [width=0.48\textwidth]{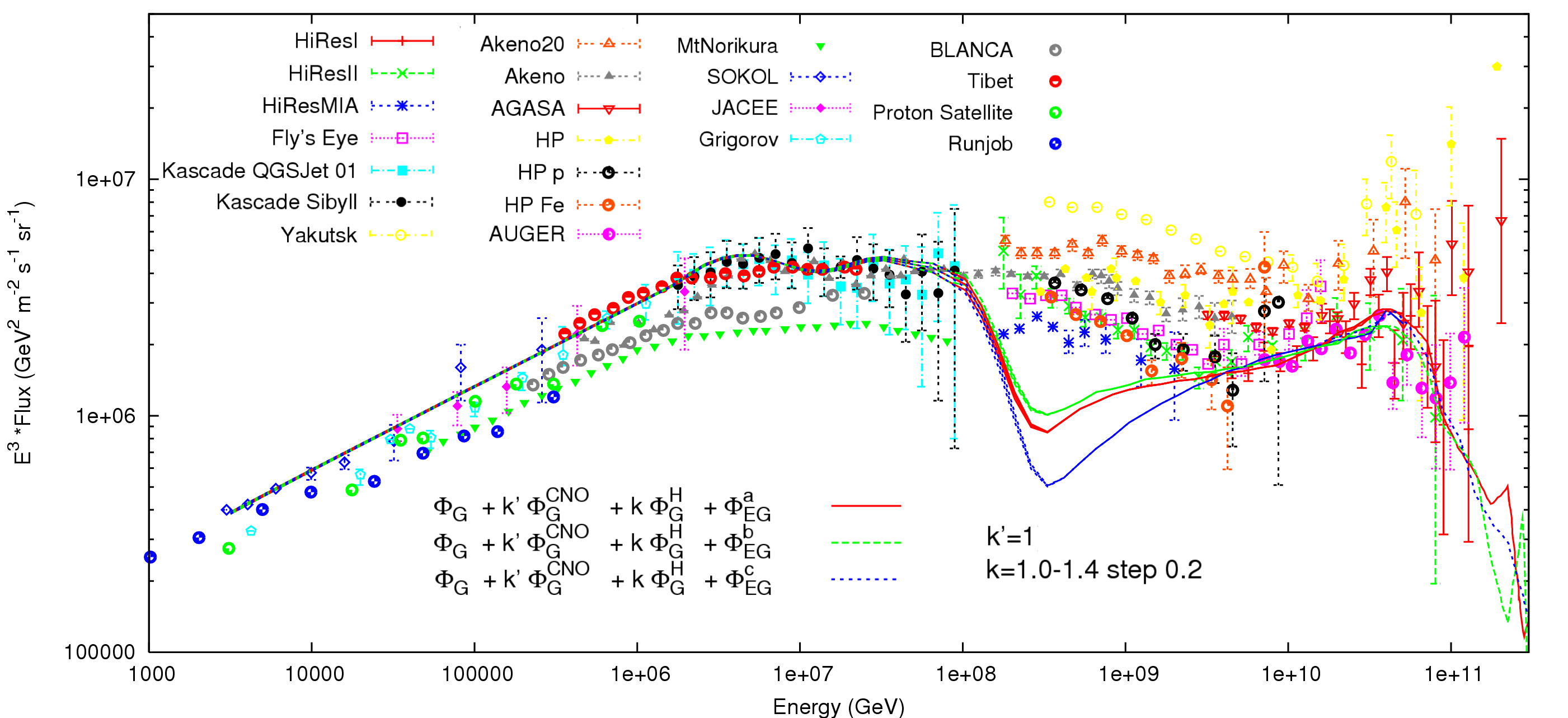}
\end{center}
\caption{Idem to Fig.\ref{BeGRLL} but for the Galactive and
Extragalactic spectrum for the {\it mixed-composition\/} model. The
CNO group ($\Phi_G^{CNO}$) of the diffusive Galactic spectrum
($\Phi_G$) has been renormalized by a factor $2$. Three source
evolution models are considered \cite{Allard2007}: (a) strong, (b)
SFR and (c) uniform. }\label{AllGR}
\end{figure}

First, we try to match the experimental data by varying the
normalization of the heavy Galactic component, while keeping
constant the previous renormalization of the CNO group. The best
reproduced spectrum for the two extragalactic models are shown in
Figs.\ref{BeGR}, \ref{BeGRLL} and \ref{AllGR}. In the case of the
{\it proton\/} model, a discontinuity appears when the two spectra
are added, regardless of the lower limit adopted for the
extragalactic component: $10^{8}$~GeV (Fig.\ref{BeGR}) or $5 \times
10^7$~GeV (Fig.\ref{BeGRLL}). The latter corresponds to cosmic
accelerators operating for the entire Hubble time.

For both, {\it proton\/} and {\it mixed-composition\/} models, there
is a flux deficit above $10^8$~GeV. The problem is much stronger for
the {\it mixed-composition\/} model where, regardless of the
luminosity evolution of EG CR sources, the total spectrum presents a
large deficit of flux between $10^8$ and $\approx 3 \times
10^9$~GeV (Fig.\ref{AllGR}).

\begin{figure}
\begin{center}
\includegraphics [width=0.48\textwidth]{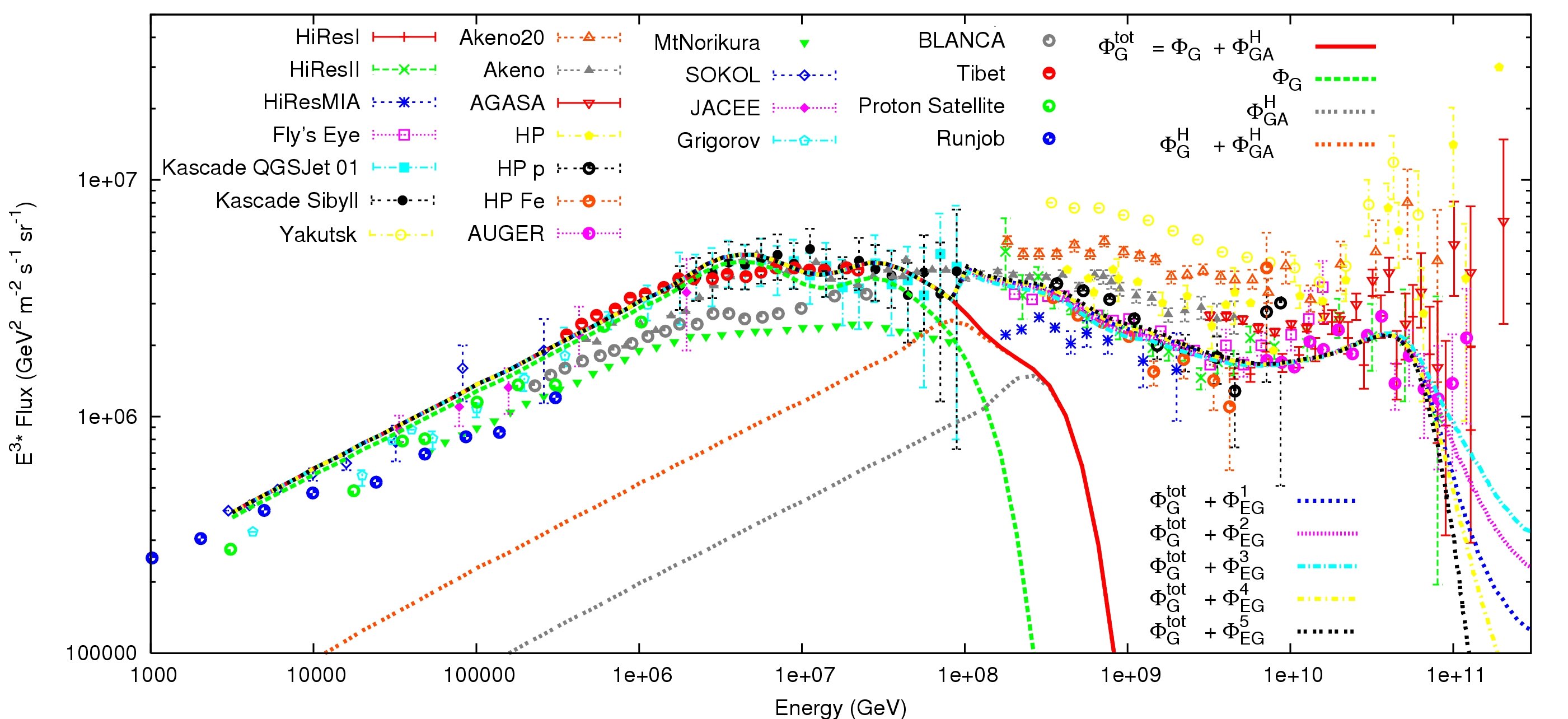}
\end{center}
\caption{Galactive and Extragalactic spectrum for the {\it proton\/}
model for an EG lower limit of $10^8~GeV$: the additional heavy
component $\phi_{GA}^H$ and the total heavy component ($\phi_{G}^H +
\phi_{GA}^H$) are also shown. }\label{BeGAc}
\end{figure}

\begin{figure}[t]
\begin{center}
\includegraphics [width=0.48\textwidth]{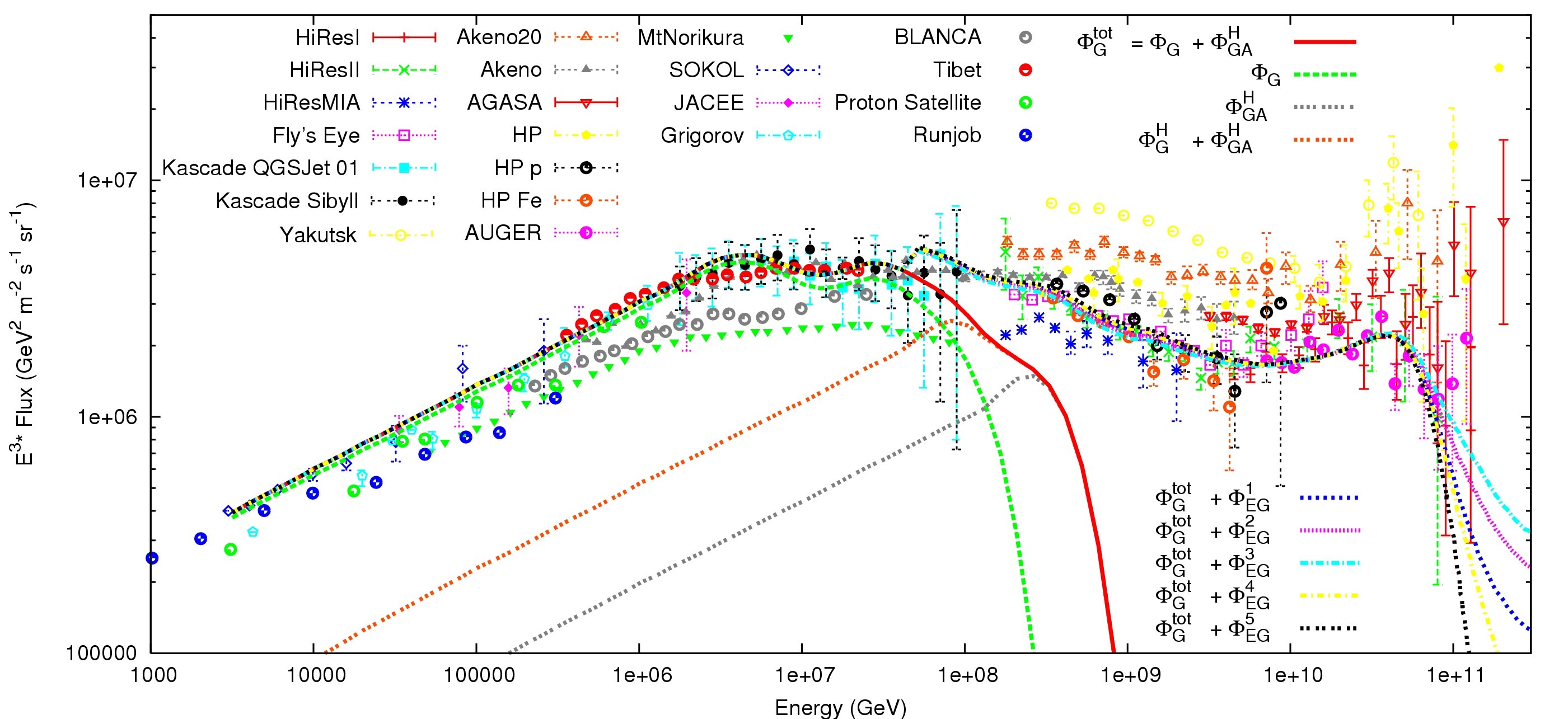}
\end{center}
\caption{Idem to Fig.\ref{BeGAc} but for an EG lower limit of $5
\times 10^7~GeV$: the additional heavy component $\phi_{GA}^H$ and
the total heavy component ($\phi_{G}^H + \phi_{GA}^H$) are also
shown.}\label{BeGALLc}
\end{figure}

\begin{figure}[t]
\begin{center}
\includegraphics [width=0.48\textwidth]{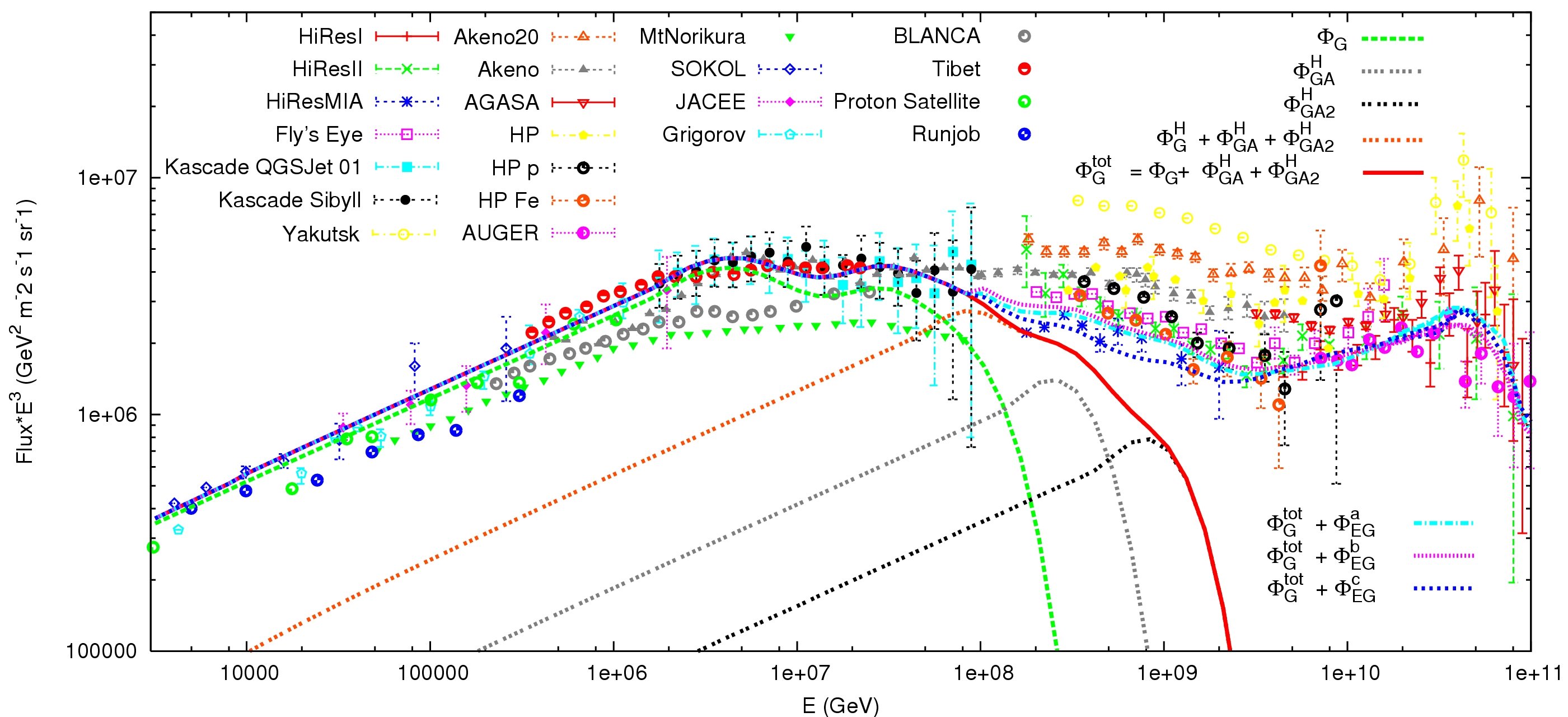}
\end{center}
\caption{Galactive and Extragalactic spectrum for the {\it
mixed-composition\/} model: the two additional heavy components
$\phi_{GA}^H$, $\phi_{GA2}^H$ and the total heavy component
($\phi_{G}^H + \phi_{GA}^H + \phi_{GA2}^H$) are shown.}
\label{AlGAc}
\end{figure}

In order to solve this flux deficit, the only way out seems to be
the introduction of an additional Galactic component. We estimate
this component by subtracting the sum of the diffusive Galactic and
extragalactic fluxes from a smooth fit to the world data. This
method confirms us the need of the renormalization of the CNO group
by a factor $\approx 2.2$ and $\approx 2$ for the {\it proton\/} and
{\it mixed-composition\/} models, respectively. In the case of the
{\it proton\/} models (Figs.\ref{BeGAc}, \ref{BeGALLc}), the
observed deficit can be resolved with an additional heavy component,
while in the {\it mixed-composition\/} models this is not enough and
we need one more additional heavy component (Fig.\ref{AlGAc}). The
additional component common to both families of models is obtained
with a shift in energy of a factor $\approx 1.5$ of the diffusive
galactic heavy component, renormalized by a factor $\approx 0.04$ in
the case of the {\it mixed-composition\/} models and $\lesssim 0.03$
for the {\it proton\/} models respectively. The second additional
component is obtained in an analogous way but with an energy-shift
factor of $\approx 1.8$ and a renormalization  by a factor $5 \times
10^{-4}$. The corresponding spectra are shown in Figs.\ref{BeGAc},
\ref{BeGALLc} and \ref{AlGAc}.

\section{Discussion}

We have analyzed the matching conditions of the Galactic and
extragalactic components of cosmic rays along the second knee and
the ankle.

It seems clear that an acceptable matching of the Galactic and
extragalactic fluxes can only be achieved if the Galaxy has
additional accelerators, besides the regular SNR, operating in the
interstellar medium. In the particular case of the {\it proton\/}
model, only one additional component is required, and this could
well represent the contribution from compact and highly magnetized
SNR, like those occurring in the central,  high density regions of
the Galactic bulge or the dense cores of molecular clouds. It must
be noted, however, that a perfectly smooth match seems unrealistic
and that some discontinuity, whose magnitude depends mainly on the
time depth from which EG CR are able to arrive at our Galaxy, should
be eventually observable in the combined spectrum.

The matching of the {\it mixed-composition\/} model is more
complicated. The Galactic spectrum has to be extended up to the
middle of the dip and this requires, besides the previous additional
component, another high energy Galactic component. The origin of
these cosmic rays pushes even further the acceleration requirements
imposed on the Galaxy and probably rapidly spinning inductors, like
neutron stars, could be invoked to fill in the gap. If this were the
case, it is very likely that photon emission at TeV energies should
uncover the sources. Changes in propagation regime inside the Galaxy at these high energies
should manifest  as a dipolar anisotropy if enough statistics
were available.

\section{Acknowledgements}

CDD thanks ICN-UNAM for hosting a long stay and Universit\`a degli Studi di Milano for a PhD
grant. GMT thanks PAPIIT/CIC-UNAM for support.

\bibliography{icrc1249}
\bibliographystyle{elsart-num}

\end{document}